%% file: main.tex
\documentclass[10pt,journal]{IEEEtran}

\input{setup/preamble.tex}
\input{setup/info.tex}

\begin{document}

\maketitle
\input{1_abstract.tex}
\input{2_introduction.tex}
\input{3_system_model.tex}
\input{4_proposed_solution.tex}
\input{5_simulation_results.tex}

\input{6_conclusion.tex}
\input{7_appendix.tex}

\balance
\bibliographystyle{IEEEtran}
\bibliography{reference/mybib}

\end{document}

%% file: setup/preamble.tex
\usepackage{amssymb}
\usepackage{amsmath}
\usepackage{cite}
\usepackage{url}
\usepackage{xcolor}
\usepackage{cite,graphicx,amsmath,amssymb}
\usepackage{fancyhdr}
\usepackage{mdwmath}
\usepackage{mdwtab}
\usepackage{caption}
\usepackage{amsthm}
\usepackage{setspace}
\usepackage{hyperref}
\usepackage{algorithm}
\usepackage{algorithmic}
\usepackage{pdfpages}
\usepackage{mathtools}
\usepackage{subcaption}
\usepackage{balance}
\hypersetup{colorlinks=true,
linkcolor=blue,
citecolor=blue,      
urlcolor=black,
}

\graphicspath{{./src/img/}}

\newtheorem{remark}{Remark}
\newtheorem{theorem}{Theorem}

\newtheorem{lemma}{Lemma}

\newtheorem{corollary}{Corollary}

\allowdisplaybreaks

%% file: setup/info.tex
\title{Near-Field Integrated Sensing and Communications}
\author{

        Zhaolin~Wang,~\IEEEmembership{Graduate Student Member,~IEEE,}
        Xidong~Mu,~\IEEEmembership{Member,~IEEE,} \\
        and Yuanwei~Liu,~\IEEEmembership{Senior Member,~IEEE}
\thanks{The authors are with the School of Electronic Engineering
and Computer Science, Queen Mary University of London, London E1 4NS,
U.K. (e-mail: zhaolin.wang@qmul.ac.uk, xidong.mu@qmul.ac.uk, yuanwei.liu@qmul.ac.uk).}
}

%% file: 1_abstract.tex
\begin{abstract}
    A near-field integrated sensing and communications (ISAC) framework is proposed, which introduces an additional \emph{distance} dimension for both sensing and communications compared to the conventional far-field system. In particular, the Cram{\'e}r-Rao bound for the near-field \emph{joint distance and angle} sensing is derived, which is minimized subject to the minimum communication rate requirement of each user. Both fully digital antennas and hybrid digital and analog antennas are investigated. For fully digital antennas, a globally optimal solution of the ISAC waveform is obtained via semidefinite relaxation. For hybrid antennas, a high-quality solution is obtained through two-stage optimization. Numerical results demonstrate the performance gain introduced by the additional distance dimension of the near-field ISAC over the far-field ISAC.
\end{abstract}

%% file: 2_introduction.tex
\section{Introduction}
Integrated sensing and communications (ISAC) has been regarded as a promising technique for the sixth-generation (6G) wireless network, where the sensing function and the communication function can be carried out simultaneously by sharing the same spectrum and hardware facilities \cite{liu2022integrated}. To fulfill the increasing demand for communication and sensing performance, future ISAC systems will evolve towards extremely large-scale antenna arrays (ELAAs) and high frequencies, e.g., millimeter wave (mmWave) and terahertz (THz), which are essential for high communication capacity and high sensing resolution \cite{liu2022integrated, mu2023noma}. Nevertheless, such a trend will significantly change the electromagnetic properties of the wireless environment, i.e., from planar-wave propagation to spherical-wave propagation, leading to an inevitable near-field effect \cite{selvan2017fraunhofer, bjornson2019massive, liu2023near}. Thus, there can be a mismatch between existing ISAC designs relying on the far-field assumption \cite{liu2020joint_signal, liu2021cramer, wang2023stars} and real wireless environments, which requires a redesign of ISAC.

Furthermore, the near-field effect also provides new possibilities. The spherical wave propagation in the near field introduces a new \emph{distance} dimension, which has the potential to facilitate joint estimation of distance and angle \cite{huang1991near} in sensing and to mitigate interference in communications \cite{zhang2022beam}. To the best of the authors' knowledge, near-field ISAC systems have not been studied yet, which motivates this work. In this letter, we propose a near-field ISAC framework. Based on the spherical wave propagation, we establish the accurate near-field channel model of the near-field ISAC framework. We jointly optimize the ISAC signal to maximize the near-field sensing performance subject to the minimum near-field communication rate for both fully digital antennas and hybrid digital and analog antennas. Finally, our numerical results verify the effectiveness of the proposed framework.

%% file: 3_system_model.tex
\section{System Model} \label{sec:system_model}

\begin{figure}[t!]
    \centering
    \includegraphics[width=0.3\textwidth]{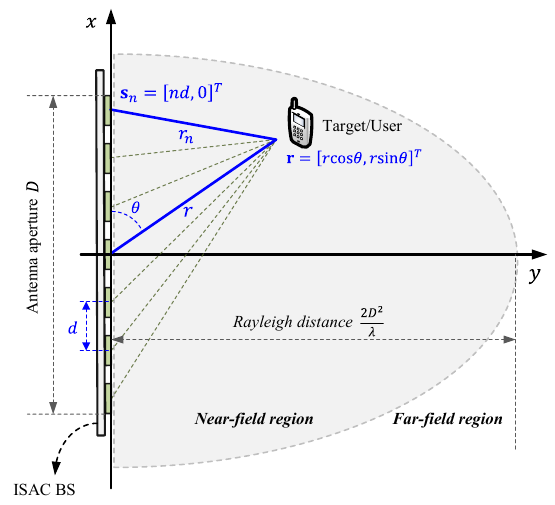}
    \caption{Illustration of the near-field ISAC system.}
    \label{fig:near_field_model}
\end{figure}

As shown in Fig. \ref{fig:near_field_model}, we consider a narrowband near-field ISAC system comprising an $N$-antenna dual-functional base station (BS), where $N = 2 \tilde{N} + 1$, $K$ single antenna communication users, and one sensing target. We consider a monostatic sensing setup at the BS. Furthermore, we assume that the BS employs a uniform linear array (ULA) with an antenna spacing of $d$, resulting in an aperture of $D = (N-1)d$. Typically, the boundary between near-field and far-field can be determined by the Rayleigh distance $\frac{2D^2}{\lambda}$ \cite{selvan2017fraunhofer}, where $\lambda$ is the signal wavelength. We assume that the communication users and sensing target are located in the near-field region of the BS, which implies that their distance from the BS is less than $\frac{2D^2}{\lambda}$. As observed, in order to take advantage of the new features of the near field, ELAAs and high frequencies are required to produce a large near-field region. Therefore, in the following, we first discuss the optimal full digital antennas to provide a performance upper bound of the near-field ISAC system. Then, we extend to the hybrid digital and analog antennas, which are more energy-efficient for ELAAs with high frequencies.

\subsection{Near-field Channel Model for ISAC}
We start with introducing the channel model for the considered near-field ISAC system. Without the loss of generality, we put the origin of the coordinate system into the center of the ULA at the BS. Therefore, the coordinate of the $n$-th element of the ULA is given by $\mathbf{s}_n = [nd, 0]^T, \forall n \in \{-\tilde{N},...,\tilde{N}\}$. Let us consider a communication user or sensing target located at a distance of $r$ and an angle of $\theta$ from the center of the ULA. Its coordinate is given by $\mathbf{r} = [r \cos\theta, r\sin\theta]^T$. Then, the distance from the $n$-th antenna element to this user or target can be calculated as follows: 
\begin{align} \label{distance}
    r_n(r, \theta) = & \|\mathbf{r} - \mathbf{s}_n\| = \sqrt{ r^2 + n^2 d^2 - 2 r n d \cos \theta  }.
\end{align}
Furthermore, in the Fresnel region of the near field, i.e., $1.2 D \le r \le \frac{2D^2}{\lambda}$, the channel gain of each link between the antenna elements and the user or target is approximately identical \cite{bjornson2021primer}. As a result, the channel gain of all links can be calculated as the free-space pathloss of the central link, which is given by $\tilde{\beta} = \frac{\sqrt{\rho_0}}{r}$, where $\rho_0 = \frac{\lambda}{4 \pi}$ is the pathloss at the reference distance $1$ m. Therefore, the channel between the $n$-th antenna element and the user or target is given by \cite{tse2005fundamentals}
\begin{equation}
    h_n(r, \theta) = \tilde{\beta} e^{-j\frac{2\pi}{\lambda} r_n(r, \theta)} = \beta e^{-j \frac{2\pi}{\lambda}(r_n(r, \theta) - r) },
\end{equation}
where $\beta = \tilde{\beta} e^{-j \frac{2\pi}{\lambda} r}$ denotes the complex channel gain. The near-field channel vector $\mathbf{h} \in \mathbb{C}^{N \times 1}$ between the BS and the user or target is given by  
\begin{align}
    \mathbf{h} = [h_{-\tilde{N}}(r_k, \theta_k),\dots, h_{\tilde{N}}(r_k, \theta_k)]^T = \beta \mathbf{a}(r, \theta),
\end{align}
where $\mathbf{a}(r, \theta)$ denotes the near-field array response vector. The $n$-th element, $\forall n \in \{ -\tilde{N},...,\tilde{N} \}$, of $\mathbf{a}(r, \theta)$ is given by   
\begin{equation} \label{eqn:beam_focus_r}
    [\mathbf{a}(r, \theta)]_n = e^{-j \frac{2\pi}{\lambda}(r_n(r, \theta) - r)}.
\end{equation}
The near-field array response vector reduces to the far-field one by assuming that $D \ll r$ and applying the first-order Taylor approximation $\sqrt{1+x} \approx 1 + \frac{1}{2}x$ for $x = \frac{1}{r^2} (n^2d^2 - 2rnd \cos\theta)$ to the distance in \eqref{distance}. In this case, the $n$-th element for the far-field array response vector is given by  
\begin{equation}
    [\mathbf{a}_{\mathrm{far}}(\theta)]_n = e^{-j \frac{2\pi}{\lambda}(-nd\cos\theta)}.
\end{equation}

Let $r_k$, $\theta_k$, and $\beta_k$ denote the distance, angle, and complex channel gain of user $k$, respectively. The near-field communication channel vector $\mathbf{h}_k \in \mathbb{C}^{N \times 1}$ between the BS and the user $k$ can be modeled as 
\begin{equation} \label{commu_channel}
    \mathbf{h}_k = \mathbf{\beta}_k \mathbf{a}(r_k, \theta_k),
\end{equation} 
In contrast to communication, target sensing based on the monostatic sensing setup relies on the echo signal received at the BS. Therefore, the round-trip channel needs to be considered. Let $r_s$, $\theta_s$, and $\beta_s$ denote the distance, angle, and complex channel gain of the sensing target, the near-field round-trip channel matrix $\mathbf{G} \in \mathbb{C}^{N \times N}$ for target sensing is given by
\begin{equation} \label{eqn:target_response}
    \mathbf{G} = \beta_s \mathbf{a}(r_s, \theta_s) \mathbf{a}^T(r_s, \theta_s),
\end{equation} 

\begin{remark}
    \textbf{(Benefits of near-field communication)} \emph{According to \eqref{commu_channel}, the near-field communication channel is determined by both distance and angle of communication users. This is fundamentally different from the far-field communication channel, which only depends on the angle. Therefore, even if the users are located in the same direction, they can still be distinguished in the distance domain, resulting in low inter-user interference.}
\end{remark}

\begin{remark} \label{remark_2}
    \textbf{(Benefits of near-field sensing)} \emph{According to \eqref{eqn:target_response}, the near-field sensing channel involves both distance and angle information. Therefore, in contrast to far-field sensing systems that can only estimate angles from the sensing channel, it is possible to carry out the joint distance and angle estimation in near-field sensing systems.}
\end{remark}

\subsection{ISAC Model for Fully Digital Antennas}
In this subsection, we consider the fully digital antenna, where each antenna at the BS is connected to a dedicated radio-frequency (RF) chain. Furthermore, we consider a coherent time block of length $T$, during which the communication channels and sensing target parameters remain approximately constant. At the beginning of each coherent time block, the communication channels are obtained by the conventional channel estimation methods. To jointly carry out communication and sensing in the remaining time of the coherent time block, the BS transmits the following joint communication and sensing signal at time $t$: 
\begin{equation} \label{t_signal}
    \mathbf{x}[t] = \sum_{k \in \mathcal{K}} \mathbf{p}_k c_k[t] + \mathbf{s}[t],
\end{equation}
where $\mathbf{p}_k \in \mathbb{C}^{N \times 1}$ denotes the fully digital beamformer for conveying the information symbol $c_k[t] \in \mathbb{C}$ to the user $k$, $\mathbf{s}[t]$ denotes the dedicated sensing signal for achieving the full sensing degrees of freedom \cite{liu2020joint_signal}, and $\mathcal{K} = \{1,\dots,K\}$. The information symbols are assumed to be independently distributed and have unit power, i.e., $\mathbb{E}\big[c_k[t] c_i^*[t]\big] = 1$, if $k=i$; and $\mathbb{E}\big[c_k[t] c_i^*[t]\big] = 0$, otherwise. Let $\mathbf{R}_s = \mathbb{E}\big[\mathbf{s}[t] \mathbf{s}^H[t]\big]$ denote the covariance matrix of the dedicated sensing signal. Then, the covariance matrix of the sensing signal is given by 
\begin{equation}
    \mathbf{R}_x = \mathbb{E}\big[\mathbf{x}[t] \mathbf{x}^H[t]\big] = \sum_{k \in \mathcal{K}} \mathbf{p}_k \mathbf{p}_k^H + \mathbf{R}_s.
\end{equation} 

\subsubsection{Communication Model}
The received communication signal at user $k$ is given by 
\begin{equation}
    y_k[t] = \underbrace{\mathbf{h}_k^T \mathbf{p}_k c_k[t]}_{\text{desired signal}} + \sum_{i \neq k} \mathbf{h}_k^T \mathbf{p}_i c_i[t] + \mathbf{h}_k^T \mathbf{s}[t] + z_k[t],
\end{equation} 
where $z_k[t] \sim \mathcal{CN}(0, \sigma_k^2)$ denotes the additive white Gaussian noise (AWGN).
Then, the achievable communication rate of user $k$ is given by \cite{tse2005fundamentals}
\begin{equation} \label{rate}
    R(\mathbf{p}_k, \mathbf{R}_x) = \log_2 \left(1 +  \frac{ |\mathbf{h}_k^T \mathbf{p}_k|^2 }{\mathbf{h}_k^T \mathbf{R}_x \mathbf{h}_k^* -  |\mathbf{h}_k^T \mathbf{p}_k|^2 + \sigma_k^2} \right).
\end{equation}  

\subsubsection{Sensing Model}
The received echo signal at the BS for target sensing is given by 
\begin{equation}
    \mathbf{y}_s[t] = \mathbf{G}\mathbf{x}[t] + \mathbf{z}_s[t],
\end{equation}
where $\mathbf{z}_s[t] \sim \mathcal{CN}(\mathbf{0}_N, \sigma_s^2 \mathbf{I}_N)$ denotes the AWGN. The objective of sensing is to estimate the target parameters from the received echo signal samples over the whole coherent time block, i.e., $\mathbf{Y}_s = [\mathbf{y}_s[1],\dots,\mathbf{y}_s[T]]$. As analyzed in Remark \ref{remark_2}, the joint distance and angle estimation can be carried out based on the near-field sensing channel $\mathbf{G}$. To this end, the concept of the classic MUltiple SIgnal Classification (MUSIC) algorithm can be invoked, where the orthogonality between the signal subspaces is exploited. The details of using the MUSIC algorithm to jointly estimate distance and angle are given in Appendix A. Typically, the mean square errors (MSE) between the estimated $(\hat{r}_s, \hat{\theta}_s)$ and the real $(r_s, \theta_s)$ is used to evaluate the sensing performance, i.e., $\epsilon_{r_s}^2 = \mathbb{E}[ |r_s - \hat{r}_s|^2 ]$ and $\epsilon_{\theta_s}^2 = \mathbb{E}[ |\theta_s - \hat{\theta}_s|^2 ]$. However, it is difficult to obtain the closed-form expression of the MSEs $\epsilon_{r_s}^2$ and $\epsilon_{\theta_s}^2$. As a remedy, we adopt the Cram{\'e}r-Rao bound (CRB) as the performance matrix for target sensing, which provides a lower bound of the MSE and has a closed-form expression \cite{kay1993fundamentals}. The CRB matrix is given by\footnote{For the cases with $M$ targets, the CRB matrix would be in the same form as \eqref{crb}, but has a larger dimension of $2M \times 2M$. The resulting optimization problem can also be solved by the algorithm proposed in Section \ref{solution}.}
\begin{equation} \label{crb}
    \mathrm{CRB}(\mathbf{R}_x, \mathbf{G}, \sigma_s^2) = \left(\mathbf{J}_{11} - \mathbf{J}_{12} \mathbf{J}_{22}^{-1} \mathbf{J}_{12}^T \right)^{-1},
\end{equation}
where the exact expressions of $\mathbf{J}_{11}$, $\mathbf{J}_{12}$, and $\mathbf{J}_{22}$ and the detailed derivation of the CRB matrix are given in Appendix A. More particularly, we have $\epsilon_{r_s}^2 \ge [\mathrm{CRB}]_{1,1}$ and $\epsilon_{\theta_s}^2 \ge [\mathrm{CRB}]_{2,2}$.

\subsection{ISAC Model for Hybrid Digital and Analog Antennas}
In this subsection, we consider the hybrid digital and analog antennas, which consist of a large-dimensional analog component realized by power-efficient phase shifters (PSs) and a low-dimensional digital component \cite{heath2016overview}. We assume that there are $N_{\mathrm{RF}}$ RF chains in the digital component, where $K+1 \le N_{\mathrm{RF}} \ll N$. Then, the transmit joint communication and sensing signal using the hybrid digital and analog antennas is given by  
\begin{equation}
    \mathbf{x}_{\mathrm{HB}}[t] = \mathbf{P}_{\mathrm{RF}} \left( \sum_{k \in \mathcal{K}} \mathbf{p}_{\mathrm{BB},k} c_k[t] + \mathbf{s}_{\mathrm{BB}}[t] \right),
\end{equation}
where $\mathbf{P}_{\mathrm{RF}} \in \mathbb{C}^{N \times N_{\mathrm{RF}}}$ denotes the analog beamformer realized by PSs, $\mathbf{p}_{\mathrm{BB},k} \in \mathbb{C}^{N_{\mathrm{RF}} \times 1}$ denotes the digital beamformer for user $k$, and $\mathbf{s}_{\mathrm{BB}}[t] \in \mathbb{C}^{N_{\mathrm{RF}} \times 1}$ denotes the digital dedicated sensing signal, whose covariance is denoted by $\mathbf{R}_{\mathrm{BB},s}$. The covariance matrix of the transmit signal is given by $\mathbf{R}_{\mathrm{HB}} = \mathbf{P}_{\mathrm{RF}} \mathbf{R}_{\mathrm{BB}} \mathbf{P}_{\mathrm{RF}}^H$, where $\mathbf{R}_{\mathrm{BB}} =  \sum_{k \in \mathcal{K}} \mathbf{p}_{\mathrm{BB},k} \mathbf{p}_{\mathrm{BB},k}^H + \mathbf{R}_{\mathrm{BB},s} $.
Furthermore, due to the hardware limitation of PSs, the analog beamformer should satisfy a unit-modulus constraint, i.e., $|[\mathbf{P}_{\mathrm{RF}}]_{m,n}| = 1, \forall m,n$. 

\subsubsection{Communication Model}
The achievable rate achieved by the hybrid digital and analog antenna for user $k$ can be directly obtained according to \eqref{rate}, which is given by $R(\mathbf{p}_{\mathrm{HB},k}, \mathbf{R}_{\mathrm{HB}})$, where $\mathbf{p}_{\mathrm{HB},k} = \mathbf{P}_{\mathrm{RF}} \mathbf{p}_{\mathrm{BB},k}$.  

\subsubsection{Sensing Model}
To receive the echo signal through the hybrid digital and analog antenna, an analog combiner $\mathbf{W}_{\mathrm{RF}} \in \mathbb{C}^{N_{\mathrm{RF}} \times N}$ should be used \cite{liu2020joint}, which is also subject to a unit-modulus constraint. Then, the received sensing signal becomes 
\begin{equation}
    \mathbf{y}_{\mathrm{HB}}^s[t] = \mathbf{W}_{\mathrm{RF}} \mathbf{G} \mathbf{x}_{\mathrm{HB}}[t] + \mathbf{W}_{\mathrm{RF}} \mathbf{z}_s[t]. 
\end{equation}
As suggested in \cite{liu2020joint}, the analog combination $\mathbf{W}_{\mathrm{RF}}$ can be randomly selected from the unit circle for target sensing. Therefore, when the number of receive antennas is
sufficiently large, it holds that $\frac{1}{N} \mathbf{W}_{\mathrm{RF}} \mathbf{W}_{\mathrm{RF}}^H \approx \mathbf{I}_{N_\mathrm{RF}}$. In this case, the effective noise $\mathbf{z}_{\mathrm{HB}}^s = \mathbf{W}_{\mathrm{RF}} \mathbf{z}_s[t]$ has a distribution of $\mathcal{CN}(\mathbf{0}_{N_\mathrm{RF}}, N \sigma_s^2 \mathbf{I}_{N_\mathrm{RF}})$. In this case, the new MUSIC algorithm and CRB matrix can be developed based on the effective sensing channel matrix $\mathbf{G}_{\mathrm{HB}} = \mathbf{W}_{\mathrm{RF}} \mathbf{G}$ and the effective noise $\mathbf{z}_{\mathrm{HB}}^s$ following the process in Appendixes A and B, respectively. The CRB matrix for hybrid digital and analog antennas can be expressed as $\mathrm{CRB}(\mathbf{R}_{\mathrm{HB}}, \mathbf{W}_{\mathrm{RF}}\mathbf{G}, N\sigma_s^2)$ 

%% file: 4_proposed_solution.tex
\section{Problem Formulation and Proposed Solution} \label{solution}

\subsection{Problem Formulation}
In the letter, we aim to minimize the CRBs on joint distance and angle estimation, while guaranteeing the minimum communication rate of each communication user. Note that in practice, target parameters only change slightly between neighboring coherent time blocks. Therefore, the estimation results of the distance and angle of the target in the previous coherent time block can be exploited for the system design. Therefore, we assume that the distance $r$ and angle $\theta$ of the target are fixed in the optimization problem \cite{liu2021cramer,wang2023stars}. Since only the diagonal entries of the CRB matrix are related to the estimation error, the optimization problem for the fully digital antenna can be formulated as follows:
\begin{subequations} \label{problem_FD}
    \begin{align}
        \min_{\mathbf{p}_k, \mathbf{R}_x \succeq 0} \quad & \mathrm{tr}\big(\mathrm{CRB}(\mathbf{R}_x, \mathbf{G}, \sigma_s^2)\big), \\
        \label{rate_FD}
        \mathrm{s.t.} \quad & R(\mathbf{p}_k, \mathbf{R}_x) \ge R_{\min,k}, \forall k, \\  
        \label{power_FD}      
        &\mathrm{tr}(\mathbf{R}_x) \le P_{\max}, \\
        \label{semi_FD}   
        & \mathbf{R}_x \succeq \sum_{k \in \mathcal{K}} \mathbf{p}_k \mathbf{p}_k^H,
    \end{align}
\end{subequations}
where $R_{\min,k}$ denotes the minimum rate requirement of user $k$ and $P_{\max}$ denotes the maximum transmit power. The last constraint \eqref{semi_FD} is from the condition that $\mathbf{R}_s = \mathbf{R}_x - \sum_{k \in \mathcal{K}} \mathbf{p}_k \mathbf{p}_k^H \succeq 0$. The optimization problem for hybrid digital and analog antennas can be formulated similarly.

\subsection{Proposed Solution for Fully Digital Antennas} \label{FD_alg}
We first consider the optimization problem \eqref{problem_FD} for fully digital antennas.
One of the main obstacles to solving problem \eqref{problem_FD} is the complex form of the CRB matrix. 
To solve this issue, we first transform problem \eqref{problem_FD} into the following equivalent but more tractable form \cite[Proposition 1]{wang2023stars}:
\begin{subequations}
    \begin{align}
        \min_{\mathbf{p}_k, \mathbf{R}_x \succeq 0, \mathbf{U} \succeq 0} \quad & \mathrm{tr}(\mathbf{U}^{-1}), \\
        \mathrm{s.t.} \quad & 
        \label{crb_matrix_FD}
        \begin{bmatrix}
            \mathbf{J}_{11} - \mathbf{U} & \mathbf{J}_{12} \\
            \mathbf{J}_{12}^T & \mathbf{J}_{22}
        \end{bmatrix} \succeq 0, \\
        & \eqref{rate_FD}-\eqref{semi_FD},
    \end{align}
\end{subequations}
where $\mathbf{U} \in \mathbb{C}^{2 \times 2}$ is an auxiliary matrix. In this case, the non-convex objective function of the CRB matrix is transformed into the convex constraint \eqref{crb_matrix_FD}.
To address the non-convex constraints \eqref{rate_FD} and \eqref{semi_FD}, the semidefinite relaxation (SDR) is adopted \cite{luo2010semidefinite}. In particular, define the auxiliary variables $\mathbf{P}_k = \mathbf{p}_k \mathbf{p}_k^H$, which satisfies $\mathbf{P}_k \succeq 0$ and $\mathrm{rank}(\mathbf{P}_k) = 1$. Then, constraint \eqref{rate_FD} can be transformed into the following convex form:
\begin{align} \label{SDR_FD_1}
    \gamma_k\mathbf{h}_k^T \mathbf{P}_k \mathbf{h}_k^* \ge \mathbf{h}_k^T \mathbf{R}_x \mathbf{h}_k^* + \sigma_k^2,
\end{align}
where $\gamma_k = 1 + \frac{1}{2^{R_{\min,k}}-1}$. Constraint \eqref{rate_FD} can also be transformed into a convex form as follows:
\begin{equation} \label{SDR_FD_2}
    \mathbf{R}_{x} \succeq \sum_{k \in \mathcal{K}} \mathbf{P}_k.
\end{equation}
By omitting the rank-one constraint of $\mathbf{P}_k$, the following optimization problem can be obtained:
\begin{subequations} \label{SDR_FD}
    \begin{align}
        \min_{\mathbf{P}_k \succeq 0, \mathbf{R}_x \succeq 0, \mathbf{U} \succeq 0} \quad & \mathrm{tr}(\mathbf{U}^{-1}), \\
        \mathrm{s.t.} \quad & \eqref{power_FD}, \eqref{crb_matrix_FD}, \eqref{SDR_FD_1}, \eqref{SDR_FD_2},
    \end{align}
\end{subequations}
which is convex and can be effectively solved by the standard interior-point algorithm. Although the rank-one constraint is omitted in problem \eqref{SDR_FD}, given any globally optimal solution $\widetilde{\mathbf{P}}_k$ and $\widetilde{\mathbf{R}}_x$ to problem \eqref{SDR_FD}, we can always construct the following rank-one solution that achieves the same objective value \cite[Theorem 1]{liu2020joint_signal}:
\begin{equation}
    \mathbf{p}_k^\star = (\mathbf{h}_k^T \widetilde{\mathbf{P}}_k \mathbf{h}_k^* )^{-\frac{1}{2}} \widetilde{\mathbf{P}}_k \mathbf{h}_k^*, \quad  \mathbf{R}_x^\star = \widetilde{\mathbf{R}}_x. 
\end{equation}
Therefore, the above solution must be the globally optimal solution to problem \eqref{problem_FD}. With this globally optimal solution, the considered fully digital antenna can provide a theoretical performance upper bound for hybrid digital and analog antennas.

\subsection{Proposed Solution for Hybrid Digital and Analog Antennas} \label{HB_alg}
We continue to study the solution to the optimization problem for hybrid digital and analog antennas. 
To solve it, we consider a heuristic two-stage optimization framework proposed in \cite{alkhateeb2015limited}. In this framework, the analog beamformer is designed to maximize the array gain at the communication users and the sensing target. Then, the digital beamformers and the digital dedicated sensing signal are optimized with the designed analog beamformer. Let $\mathbf{p}_{\mathrm{RF},\ell}$ denote the $\ell$-th column of $\mathbf{P}_{\mathrm{RF}}$. Then, the analog beamformer can be designed as follows:
\begin{equation}
    \mathbf{p}_{\mathrm{RF},\ell} = \begin{cases}
        \mathbf{a}^*(r_\ell, \theta_\ell), & 1 \le \ell \le K, \\
        \mathbf{a}^*(r_s, \theta_s), & K < \ell \le N_{\mathrm{RF}}.
    \end{cases}
\end{equation}
With the above solution of the analog beamformer, the optimization problem is only related to the digital part $\mathbf{p}_{\mathrm{BB},k}$ and $\mathbf{R}_{\mathrm{BB}}$, which has the same form as the problem \eqref{problem_FD}. Therefore, it can be optimally solved by the algorithm proposed in the previous section.

%% file: 5_simulation_results.tex
\begin{figure}[t!]
  \centering
  \includegraphics[width=0.38\textwidth]{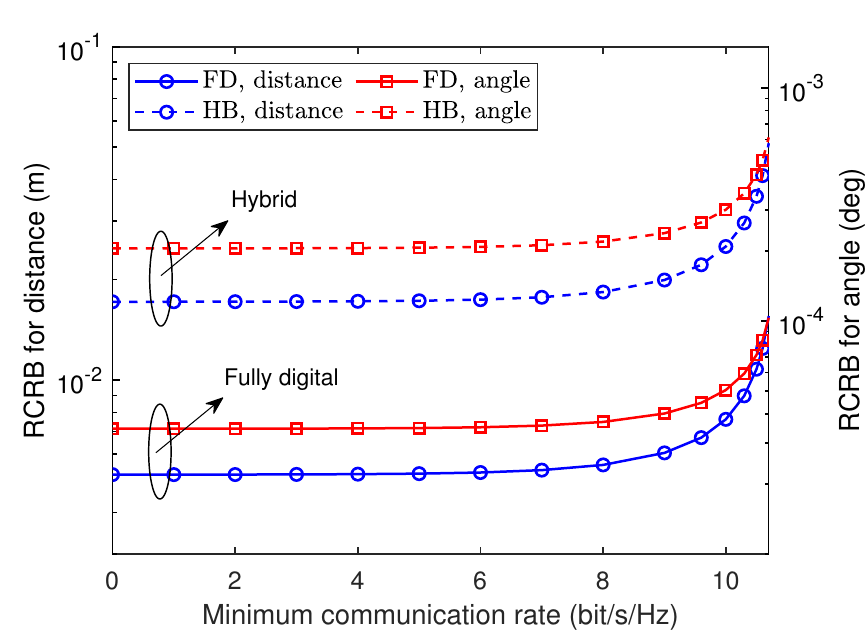}
  \caption{RCRB versus minimum communication rate.}
  \label{RCRB_vs_rate}
\end{figure}

\section{Numerical Results}

In this section, numerical results are provided to verify the effectiveness of the proposed near-field ISAC framework. We assume a BS equipped with a ULA with $N=65$ antennas operating at a frequency of $28$ GHz ($\lambda = 1.07$ cm) \cite{zhang2022beam}. The antenna aperture is set to $D = 0.5$ m, resulting in a Rayleigh distance of $\frac{2D^2}{\lambda} = 46.73$ m. There are $K = 4$ communication and one sensing target located in the near-field region of the BS. The locations of the users are randomly generated within the near-field region of the BD. The location of the target is set to $(20 \text{ m}, 45^\circ)$. 
The maximum transmit power at the BS and the noise power are set to $20$ dBm and $-60$ dBm, respectively. For the hybrid digital and analog antennas, the number of RF chains is set to $N_{\mathrm{RF}} = 5$. The minimum rate requirement of each user is assumed to be the same, i.e., $R_{\min, k} = R_{\min}, \forall k$. In particular, “FD” represents fully digital antennas, “HB” represents hybrid digital and analog antennas, and “RCRB” represents the root of CRB.

\begin{figure}[t!]
  \centering
  \begin{subfigure}[t]{0.24\textwidth}
    \centering
    \includegraphics[width=1\textwidth]{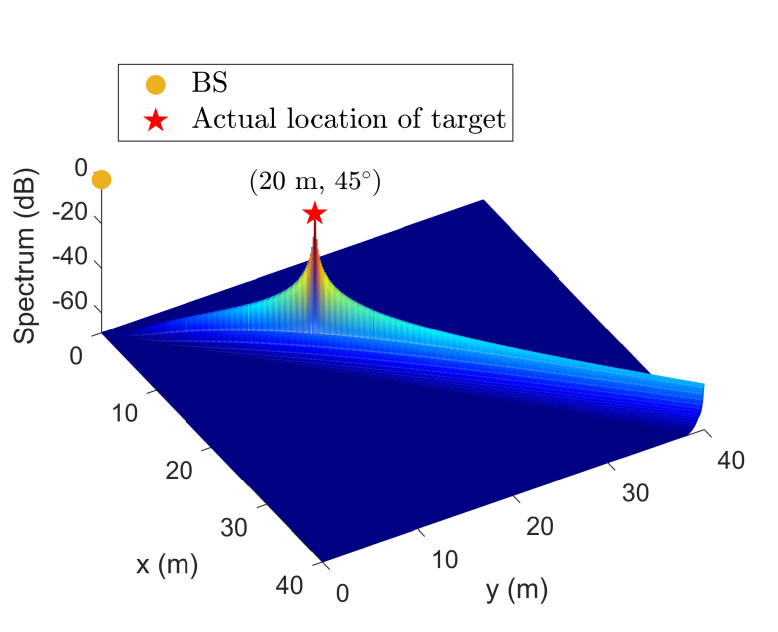}
    \caption{Near-field ISAC.}
  \end{subfigure}%
  \begin{subfigure}[t]{0.24\textwidth}
    \centering
    \includegraphics[width=1\textwidth]{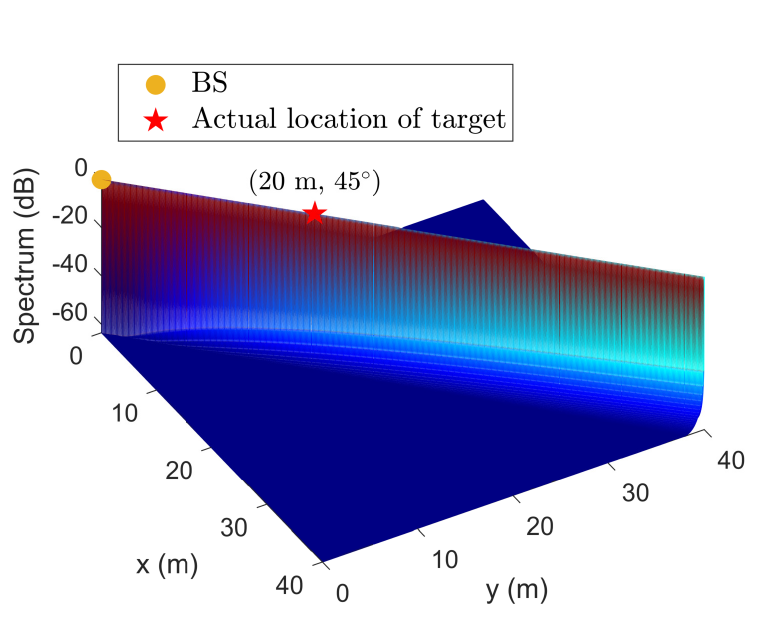}
    \caption{Far-field ISAC.}
\end{subfigure}
\caption{Normalized spectrum of MUSIC for $R_{\min} = 5$ bit/s/Hz.}
\label{fig:MUSIC}

\end{figure}

\subsection{RCRB Versus Minimum Communication Rate}
In Fig. \ref{RCRB_vs_rate}, the RCRBs for distance and angle estimation versus the minimum communication rate $R_{\min}$. As can be observed, the RCRBs increase with the increment of $R_{\min}$, indicating the existence of a tradeoff between sensing and communication performance. Nevertheless, it is noteworthy that even when $R_{\min}$ is considerably high, the RCRB remains at a low level, which confirms the effectiveness of integrating near-field sensing and near-field communications. Furthermore, HB reduces sensing performance compared to FD at the same communication rate, but it has much lower power consumption.

\begin{figure}[t!]
  \centering
  \includegraphics[width=0.38\textwidth]{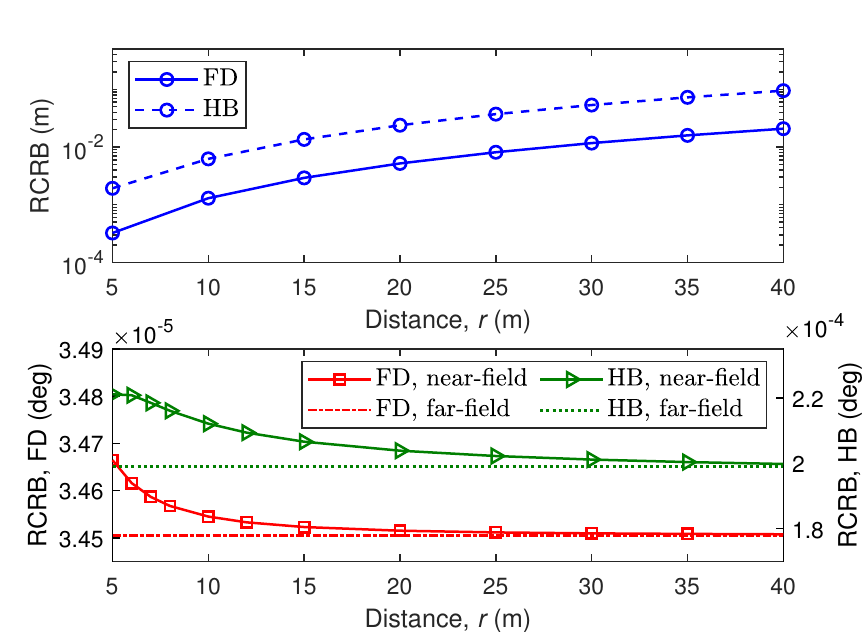}
  \caption{RCRB versus distance for $R_{\min} = 5$ bit/s/Hz. (a) Top: distance estimation; (b) Bottom: angle estimation}
\end{figure}

\subsection{Spectrum of MUSIC}
In Fig. \ref{fig:MUSIC}, we compare the normalized spectrum of MUSIC, i.e., $\frac{1}{p(r, \theta)}$, obtained by the proposed near-field ISAC and the far-field ISAC \cite{liu2021cramer} over a fine grid of $x \in [0:0.08:40]$ m and $y \in [0:0.08:40]$ m. The minimum communication rate is set to $R_{\min} = 5$ bit/s/Hz. For near-field ISAC, the largest value in the spectrum occurs around the actual location of the target. The estimated distance and angle from the spectrum of near-field ISAC is $\hat{r}_s = 19.952$ m and $\hat{\theta}_s = 45^\circ$.
However, for far-field ISAC, the spectrum has the same largest value along the direction of the target. This is because the near-field sensing channel contains information on both distances and angles, while the far-field one is only related to angles.

\subsection{RCRB Versus Distance}
To obtain more insights, we further study the impact of the distance $r$ on the near-field ISAC, without factoring in pathloss for a fair comparison. We set $R_{\min} = 5$ bit/s/Hz. As the distance $r_s$ increases, we observe a gradual increase in the RCRB of distance, indicating lower accuracy of distance estimation. This is expected because as $r_s$ increases, the near-field sensing channel gradually degrades to the far-field sensing channel and therefore holds less distance information. Furthermore, for angle estimation, it is interesting to see that the RCRB of near-field ISAC gradually decreased and is limited by the RCRB of far-field ISAC. This is also expected because, with an increase in $r_s$, the echo signal at each antenna element emanates from nearly the same direction, thereby favoring angle estimation.

%% file: 6_conclusion.tex
\section{Conclusion}
A near-field ISAC framework has been proposed. It is suggested that ISAC systems can benefit more from the near field compared with the far field. Therefore, exploring ways to enlarge the near-field region of ISAC systems without altering the number of antennas and frequencies significantly is an interesting direction for future research.

%% file: 7_appendix.tex
\section*{Appendix A\\MUSIC Algorithm for Near-field Sensing}
MUSIC algorithm is a kind of super-resolution algorithm for parameter estimation, which exploits the orthogonality of signal subspaces \cite{schmidt1986multiple}. 
The signal subspace can be obtained according to the covariance matrix of the received echo signal, i.e., $\bar{\mathbf{R}} = \mathbb{E}\big[ \mathbf{y}_s[t] \mathbf{y}_s^H[t] \big] \approx \frac{1}{T} \sum_{t=1}^T \mathbf{y}_s[t] \mathbf{y}_s^H[t]$. Then, based on eigenvalue decomposition, the signal and noise subspaces can be obtained as follows:
\begin{align}
    \bar{\mathbf{R}} = \underbrace{\mathbf{E}_s \mathbf{D}_s \mathbf{E}_s^H}_{\text{signal subspace}} + \underbrace{\mathbf{E}_n \mathbf{D}_n \mathbf{E}_n^H}_{\text{noise subspace}},
\end{align} 
where $\mathbf{D}_s$ and $\mathbf{E}_s$ contain the $M$ largest eigenvalues and corresponding eigenvectors, while $\mathbf{D}_n$ and $\mathbf{E}_n$ contain the remaining eigenvalues and eigenvectors, respectively. Here, $M=1$ represents the number of targets. In particular, $\mathbf{D}_s$ and $\mathbf{D}_n$ are real-valued diagonal matrices.
According to \eqref{eqn:target_response}, the signal subspace is spanned by the vector $ \mathbf{a}(r_s, \theta_s)$. By defining the projection operator onto the noise subspace as $\mathbf{P}_{\mathbf{E}_n} = \mathbf{E}_n (\mathbf{E}_n^H \mathbf{E})^{-1} \mathbf{E}_n^H$, for any vector $\mathbf{a}(r, \theta)$, we have its projection onto the noise space as follows \cite{schmidt1986multiple}: 
\begin{align}
    p(r, \theta) = \left\| \mathbf{P}_{\mathbf{E}_n} \mathbf{a}(r, \theta)  \right\|^2 = \mathbf{a}^H(r, \theta) \mathbf{E}_n \mathbf{E}_n^H \mathbf{a}(r, \theta).
\end{align}
Based on the orthogonality between the signal subspace and the noise subspace, it holds that $p(r, \theta) \rightarrow 0$ if and only if $r = r_s$ and $\theta = \theta_s$. Thus, the estimated distance and angle of the target are given by $(\hat{r}_s, \hat{\theta}_s) = \arg \min_{(r, \theta)} p(r, \theta)$.

\section*{Appendix B\\Derivation of CRB Matrix}
The CRB matrix can be calculated by the inverse of the Fisher information matrix (FIM) with respect to the unknown parameters. In particular, in the sensing channel $\mathbf{G}$, the unknown parameters are given by $\boldsymbol{\xi} = [r_s, \theta_s, \beta_s^r, \beta_s^i]$, where $\beta_s^r = \mathrm{Re}\{ \beta_s \}$ and $\beta_s^i = \mathrm{Im}\{ \beta_s \}$. Define $\mathbf{u} = \mathrm{vec}( \mathbf{Y}_s )$.
According to \cite[Appendix 3C]{kay1993fundamentals}, the FIM $\mathbf{J}_{\boldsymbol{\xi}}$ for estimation the unknown parameter $\boldsymbol{\xi}$ from $\mathbf{u}$ is given by
\begin{equation}
    \mathbf{J}_{\boldsymbol{\xi}} = \frac{2}{\sigma_s^2} \mathrm{Re}\left\{ \frac{\partial \mathbf{u} }{\partial \boldsymbol{\xi}} \frac{\partial \mathbf{u}^H }{\partial \boldsymbol{\xi}} \right\} = 
    \begin{bmatrix}
        \mathbf{J}_{11} & \mathbf{J}_{12} \\
        \mathbf{J}_{12}^T & \mathbf{J}_{22},
    \end{bmatrix}
\end{equation}
where $\mathbf{J}_{11} = \begin{bsmallmatrix}
    J_{r_s r_s} & J_{r_s \theta_s} \\
    J_{r_s \theta_s} & J_{\theta_s \theta_s}
\end{bsmallmatrix}$,  $\mathbf{J}_{12} = \begin{bsmallmatrix}
    J_{r_s \beta_s^r} & J_{r_s \beta_s^i} \\
    J_{\theta_s \beta_s^r} & J_{\theta_s \beta_s^i}
\end{bsmallmatrix}$, and $\mathbf{J}_{22} = \begin{bsmallmatrix}
    J_{\beta_s^r \beta_s^r} & 0 \\
    0 & J_{\beta_s^i \beta_s^i}
\end{bsmallmatrix}$.
The value of each enrty is given by $J_{\ell p} = \frac{2}{\sigma_s^2} \mathrm{Re}\left\{ \frac{\partial \mathbf{u}^H }{\partial \ell} \frac{\partial \mathbf{u} }{\partial p} \right\}$. By Defining $\tilde{\mathbf{G}} = \mathbf{a}(r_s, \theta_s) \mathbf{a}^T(r_s, \theta_s)$, $\tilde{\mathbf{G}}_{r_s} = \partial \tilde{\mathbf{G}}/\partial r_s$ and $\tilde{\mathbf{G}}_{\theta_s} = \partial \tilde{\mathbf{G}}/\partial \theta_s$, and exploiting $\mathbf{R}_x \approx \frac{1}{T} \sum_{t=1}^T \mathbf{x}[t] \mathbf{x}^H[t]$, for any $\ell, p \in \{r_s, \theta_s\}$, we have 
\begin{align}
    &J_{\ell p} = \frac{2 |\beta_s|^2 T }{\sigma_s^2} \mathrm{Re}\left\{  \mathrm{tr}( \tilde{\mathbf{G}}_p \mathbf{R}_x \tilde{\mathbf{G}}_\ell^H )  \right\}, \\
    &J_{\ell \beta_s^r} = \frac{2 T}{\sigma_s^2} \mathrm{Re} \left\{ \beta_s^* \mathrm{tr}( \tilde{\mathbf{G}} \mathbf{R}_x \tilde{\mathbf{G}}_\ell^H )  \right\}, \\  
    & J_{\ell \beta_s^i} = \frac{2 T}{\sigma_s^2} \mathrm{Re} \left\{ j \beta_s^* \mathrm{tr}( \tilde{\mathbf{G}} \mathbf{R}_x \tilde{\mathbf{G}}_\ell^H )  \right\}, \\
    &J_{\beta_s^r \beta_s^r} = J_{\beta_s^i \beta_s^i} = \frac{2T}{\sigma_s^2} \mathrm{tr}\left( \tilde{\mathbf{G}} \mathbf{R}_x \tilde{\mathbf{G}}^H \right).
\end{align}
Based on the FIM, the CRB matrix for estimating $r_s$ and $\theta_s$ is given by $\mathrm{CRB}(\mathbf{R}_x, \mathbf{G}, \sigma_s^2) = (\mathbf{J}_{11} - \mathbf{J}_{12} \mathbf{J}_{22}^{-1} \mathbf{J}_{12}^T )^{-1}$ \cite{bekkerman2006target}.